\documentclass[floatfix,showkeys,superscriptaddress,showpacs,preprintnumbers,aps,twocolumn]{revtex4}

\usepackage[latin1]{inputenc}
\usepackage{epsfig}
\usepackage{amsmath}
\usepackage{amsfonts}
\usepackage{amssymb}
\usepackage{tabularx}
\usepackage{bm}

\newcommand{\eps}{\varepsilon}

\newcommand{\ITP}{Institut f{\"u}r Theoretische Physik, 
  Technische Universit{\"a}t Berlin,
  Hardenbergstra{\ss}e 36, 10623 Berlin, Germany}

\newcommand{\CHEMNITZ}{Institut f{\"u}r Physik, 
  TU Chemnitz, D--09107 Chemnitz, 
  Germany}

\newcommand{\QUEENMARY}{School of Mathematical Sciences, 
  Queen Mary / Univ.\ of London, 
  Mile End Road, 
  London E1 4NS, UK}
\begin{document}

\date{\today}
  
\title{Time--delay autosynchronization of the spatio-temporal dynamics
in resonant tunneling diodes}

\author{J. Unkelbach}
\affiliation{\ITP} 
\author{A. Amann}
\affiliation{\ITP} 
\author{W. Just}
\altaffiliation{permanent address: \QUEENMARY} 
\affiliation{\ITP} 
\affiliation{\CHEMNITZ}
\author{E. Sch{\"o}ll}
\email{schoell@physik.tu-berlin.de} 
\affiliation{\ITP} 

\begin{abstract}
  The double barrier resonant tunneling diode exhibits complex
  spatio-temporal patterns including low-dimensional chaos when
  operated in an active external circuit.  We demonstrate how
  autosynchronization by time--delayed feedback control can be used to
  select and stabilize specific current density patterns in a noninvasive way. We compare the
  efficiency of different control schemes involving feedback in either
  local spatial or global degrees of freedom. The numerically obtained
  Floquet exponents are explained by analytical results from linear
  stability analysis.
\end{abstract}

\preprint{to be submitted to PRE}



\maketitle

\section{Introduction}\label{sec1}

Since the groundbreaking work by Ott, Grebogi and Yorke \cite{OTT90},
chaos control has evolved into a central issue in nonlinear science \cite{SCH99c}.
While earlier methods of chaos control have used a rather complicated
calculation of the control force from the Poincare map, recent control
schemes based on
time--delay autosynchronization  \cite{PYR92,SOC94} are much simpler
to handle and have been applied to a number of real world problems
\cite{BIE94,PIE96,SIM96,HAL97,SUK97,PAR99,LUE01,BEN02,BET03}. 

One intriguing aspect is the possibility of noninvasive control. This refers to
the stabilization of a target state which is not changed by the control
term, and the control force vanishes once the target state has been
reached.  A natural choice for the target state are unstable periodic
orbits (UPOs), since they are dense in the chaotic attractor of
the uncontrolled system.  

While earlier work has concentrated on low-dimensional dynamic systems
described by maps or ordinary differential equations \cite{SCH93b}, the 
emphasis has recently shifted towards stabilization of spatio-temporal patterns. 
It was shown that for a generic nonlinear reaction--diffusion model
of activator--inhibitor type
with one spatial degree of freedom, different noninvasive time--delayed feedback
methods can be used to suppress chaotic behavior
\cite{FRA99,BEC02,BAB02,JUS03}, and their respective domains of control have been
compared and interpreted in terms of Floquet spectra.

In this paper we will apply time--delayed feedback control schemes to a
semiconductor nanostructure which is currently of great interest
\cite{SCH00}: the {\em double barrier resonant tunneling diode} (DBRT).  
This device is well known as an electronic oscillator, and 
complex spatio-temporal patterns of the current density have been 
reported in numerical simulations
\cite{GLA97,MEL98,FEI98,MEI00b,ROD03}, including chaotic
spatio-temporal scenarios \cite{SCH02}.
Here we propose to apply time--delay autosynchronization to stabilize
those spatio-temporal breathing and spiking patterns under a wide range 
of operating conditions, and induce stable periodic oscillations. The
paper is organized as follows. After the introduction of the DBRT model 
in Sect.\ \ref{sec:model}, we examine the dynamical
bifurcation scenarios leading to the formation of lateral current density
patterns of
the uncontrolled system in Sect.\ \ref{sec:scenarios}. We
find parameter regimes featuring chaotic breathing as well as chaotic
spiking.  In Sect.\ \ref{sec:control} time-delayed feedback
control is used for the stabilization of periodic spiking and
breathing oscillations. We compare the control performance for different
control schemes. The details of the model are given in the Appendix.

\section{The model}
\label{sec:model}

\begin{figure}
  \epsfig{file=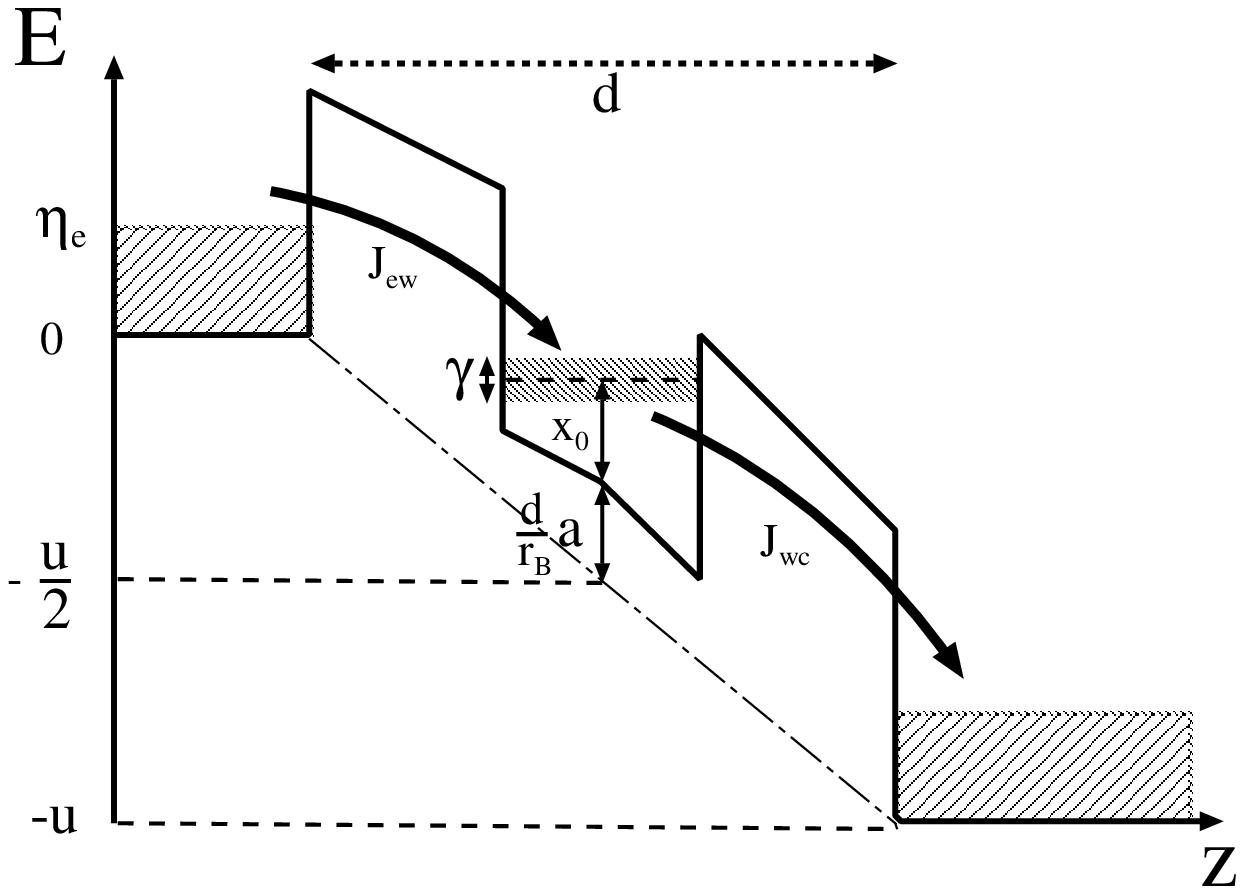,height=5cm}
  \caption{Schematic energy band structure of the DBRT. The symbols are
  explained in the Appendix.}
  \label{fig:dbrt_scheme}
\end{figure}

\begin{figure}
  \begin{center}

  \epsfig{file=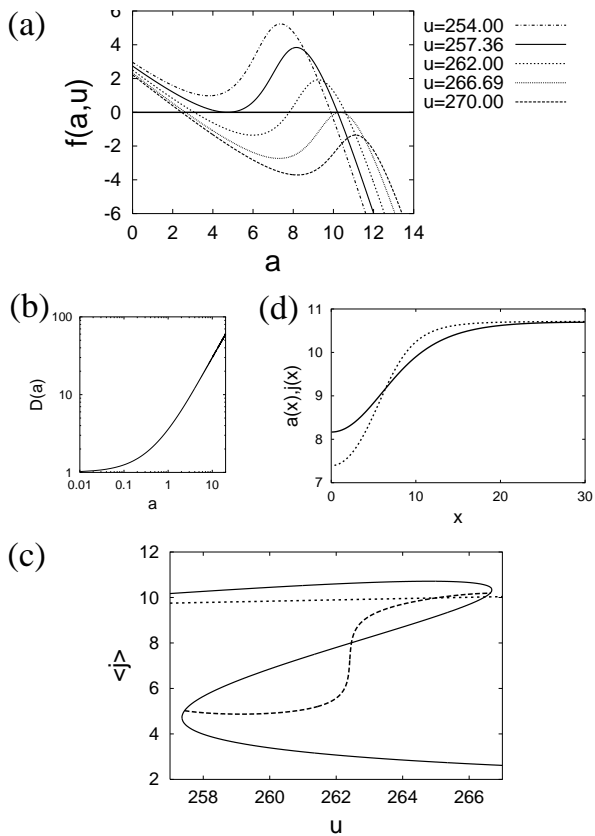,width=8cm} 
  \end{center}

  \caption{(a) Nonlinear kinetic function $f(a,u)$ for different values
    of voltage $u$, (b) diffusion coefficient
    $D(a)$ as a function of the electron density $a$ (double-logarithmic
    scale), (c) 
    current--voltage characteristic for spatially homogeneous states
    (solid line) and filamentary states (dashed line) for $L = 30$; load
    line for $U_0=-84.2895$, $r=-35$ (dotted line), (d) spatial profile
    of a half filament, $a(x)$ (solid line) and $j(x)$ (dotted line)
    ($L=30;$  $u=265$)}
  \label{fig:f_D}
\end{figure}

The DBRT is a semiconductor nanostructure which consists of one GaAs
quantum well sandwiched between two AlGaAs barriers along the $z$-direction
(cf. Fig.~\ref{fig:dbrt_scheme}). The quantum well
defines a two-dimensional electron gas in the $x-y$ plane. 
The spatially homogeneous steady states give rise to a Z-shaped 
current--voltage characteristic \cite{GOL87a} exhibiting bistability in a range 
of applied voltages. The middle branch of the current--voltage characteristic
can be stabilized by applying an appropriate active circuit with
a negative load resistance \cite{MAR94}. Complex
spatio-temporal patterns can arise if the lateral
re--distribution of electrons within the quantum well
\cite{WAC95d,GLA97,MEL98,FEI98,MEI00b,SCH02,ROD03} is taken into account.
In the following we assume that
the extension of the device in $y$-direction is small and
charge inhomogeneities can only appear in the $x$-direction.  
Using dimensionless variables throughout we arrive at the following equations:
\begin{eqnarray}
\frac{\partial a}{\partial t} &=& \frac{\partial }{\partial x} 
\left(D(a)\frac{\partial a}{\partial x}\right) + f(a,u) - KF_a(x,t)
\label{eq:a}\\
\frac{du}{dt} &=& \frac{1}{\varepsilon}
\left(U_0 - u - r \langle j \rangle\right)- KF_u(t)
\label{eq:u}
\end{eqnarray}
Here the uncontrolled model \cite{SCH02} has been extended by
the control terms $K F_a$ and $K F_u$ representing the control
forces with amplitude $K$.
The dynamic variables are the inhibitor $u(t)$ and the activator $a(x,t)$.
The one--dimensional spatial coordinate $x$ corresponds to the direction
transverse to the current flow. In
the semiconductor context $u(t)$ denotes the voltage drop across the
device and $a(x,t)$ is the electron density in the quantum well.  The
nonlinear, nonmonotonic function $f(a,u)$ describes the balance of the
incoming and outgoing current densities in the quantum well
(Fig.~\ref{fig:f_D}~(a)), and $D(a)$ is an effective, electron density
dependent transverse diffusion coefficient (Fig.~\ref{fig:f_D}~(b))
\cite{CHE00}.  The local current density in the device is
$j(a,u)=\frac{1}{2}(f(a,u)+2a)$, and $\langle j \rangle =
\frac{1}{L}\int_0^{L} j dx$ is associated with the global current. Eq.
(\ref{eq:u}) represents Kirchhoff's law of the circuit in which the
device is operated. The external bias voltage $U_0$, the 
load resistance $r$, and the time-scale ratio $\varepsilon$
are dimensionless external parameters. We consider
a system of width $L$ 
with no charge transfer through the lateral boundaries (i.e., Neumann
boundary conditions $\partial_x a=0$ at $x=0, L$).  In
Appendix~\ref{sec:app} we give explicit expressions for $f(a,u)$ and
$D(a)$ and draw the connection to the microscopic physical parameters.  

In the case without control force ($K=0$) one stationary solution of
(\ref{eq:a}) is given by the homogeneous solution
$a(x)=a_i^{\text{hom}}$ with $f(a_i^{\text{hom}},u)=0$. Up to three
different solutions $a_i^{\text{hom}}$ may exist for one fixed value of $u$
(Fig.~\ref{fig:f_D}(a)). This gives rise to a 
$Z$-shaped current--voltage characteristic $j(u)=a$ (solid
line in Fig.~\ref{fig:f_D}(c)).

The spatially homogeneous stationary solutions of the coupled equations
(\ref{eq:a}) and (\ref{eq:u}) are given by the intersection of the
current--voltage characteristic $j(u)$ and the load line which is the
nullcline $\langle j \rangle=(U_0 - u)/r$ of (\ref{eq:u}). States on the
middle branch of the $j(u)$ characteristic are unstable
in a passive external circuit with effective resistance $r>0$.
By choosing $r<0$, which can be realised by an active circuit, i.e.
applying an additional control voltage proportional to the device current
$\langle j \rangle$ in series with the bias $U_0$
\cite{MAR94}, it is in principle possible to stabilize
the middle branch of the stationary $j(u)$ characteristic and access it
experimentally, but for large enough $\eps$ Hopf bifurcations can occur,
leading to uniform limit cycle oscillations of the current and voltage. 

\begin{figure}
  \centering
  \epsfig{file=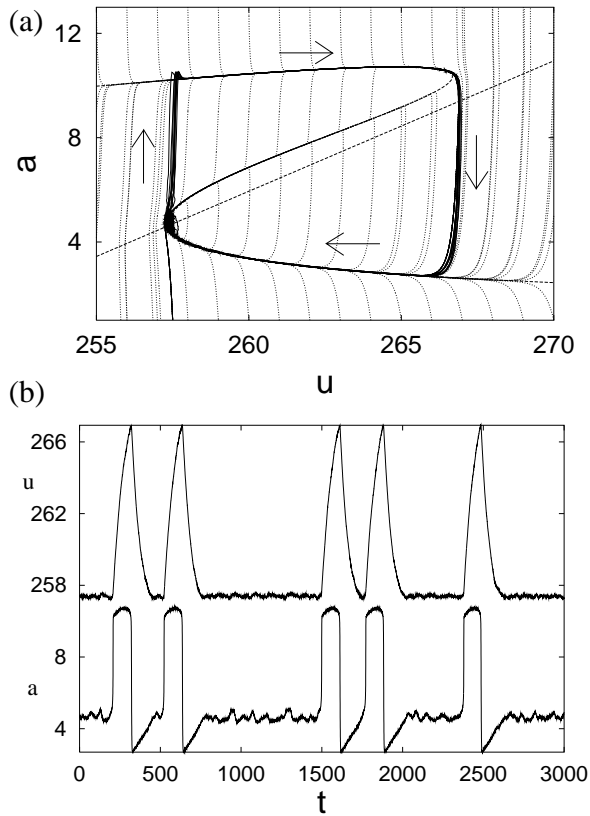,width=8cm}
  \caption{DBRT in the excitable regime:
  (a) Spatially homogeneous phase portrait of current $j\equiv a$
  vs. voltage $u$ for $r= -2$, $U_0=248.11$,
  $L=30$, $\eps =70$. Solid line: trajectory of the system under
  noise. Dashed lines: Load line (straight) and homogeneous current--voltage
  characteristic ($Z$--shaped). Dotted lines: phase flow of the system. (b)
  Response of $a$ and $u$ triggered by noise. }
  \label{fig:homogenous_scenarios}
\end{figure}

It should be noted that, depending upon the intersection of the load line
with the $j(u)$ characteristics, the DBRT represents either a bistable, or
an oscillatory, or an excitable active medium \cite{MIK94}. It is
remarkable that the DBRT can be operated as an excitable
system even in the spatially homogeneous case. Consider the situation in
Fig.~\ref{fig:homogenous_scenarios}(a) where the load line
intersects the homogeneous characteristic near the second turning
point at a low current value ($j\approx 4.63$, and $\eps=70$). Due to
the large value of $\eps$
the relaxation in $u$ is much slower than the relaxation in $a$. If
noise pushes the system off its fixed point in the low current regime,
it quickly relaxes to a value of $a\equiv j$ in a high current state. Then the slow
relaxation in $u$ sets in and drags the system along the homogeneous
characteristic towards high $u$ (since we are above the load line)
until the upper branch of the homogeneous characteristic ends. At this
point, the system again quickly relaxes in $a$, this time towards a low
current state. Since we are now below the load line the system finally
returns to the fixed point. This excursion in phase space results in a
spike-like response in the current and voltage signal upon external noise
(Fig.~\ref{fig:homogenous_scenarios}(b)) which is
characteristic for an {\em excitable} medium.

\section{Spatio-temporal scenarios}
\label{sec:scenarios}

As the device width $L$ in $x$-direction increases (the device width
in $y$ direction is fixed and small compared to $L$), the middle branch of the homogeneous
solution $a^{hom}_2$ (Fig.~\ref{fig:f_D}(c)) becomes unstable against
inhomogenous fluctuations. A straightforward linear stability
analysis \cite{SCH00} shows that this occurs at
\begin{equation}
  \label{eq:L_condition}
  L > \pi\sqrt{\left|\frac{D(a^{hom}_2)}{\partial_a f(a^{hom}_2)}\right|}.
\end{equation}
Then an additional stationary solution of eq.~(\ref{eq:a}) arises in the
form of a current filament.
Fig.~\ref{fig:f_D}(d) shows the dependence of $a$ and $j$ on the
  transverse spatial coordinate $x$ for a half--filament with $L=30$ and
  $u=265$. The dashed line in Fig.~\ref{fig:f_D}(c) depicts the corresponding
  current--voltage characteristic $\langle j \rangle(u)$. 

\begin{figure}
  \centering
  \epsfig{file=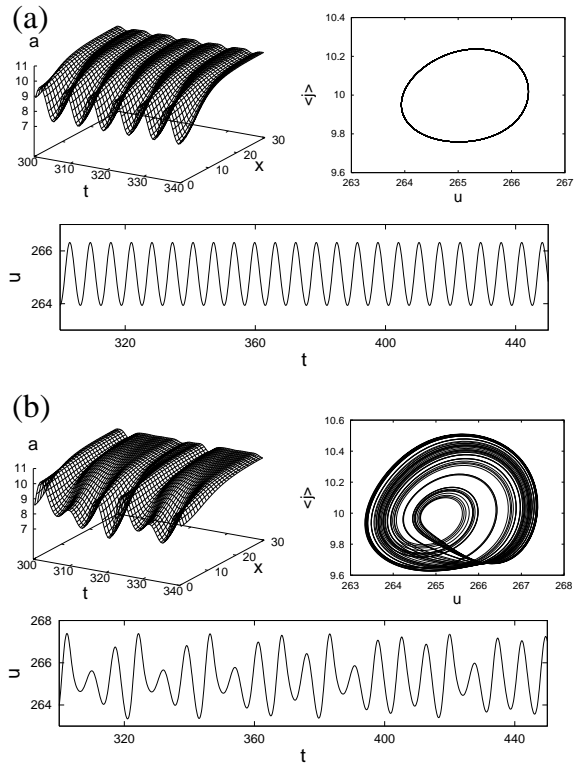,width=8cm}
  \caption{Spatio-temporal breathing patterns:
    electron density evolution, phase portrait, and voltage
    time series for (a)   $\eps=7.0$: periodic breathing,
    (b) $\eps=9.1$: chaotic breathing; load line as in Fig.~\ref{fig:f_D}}.
  \label{fig:breathing_scenarios}
\end{figure}

\begin{figure}
  \centering
    \epsfig{file=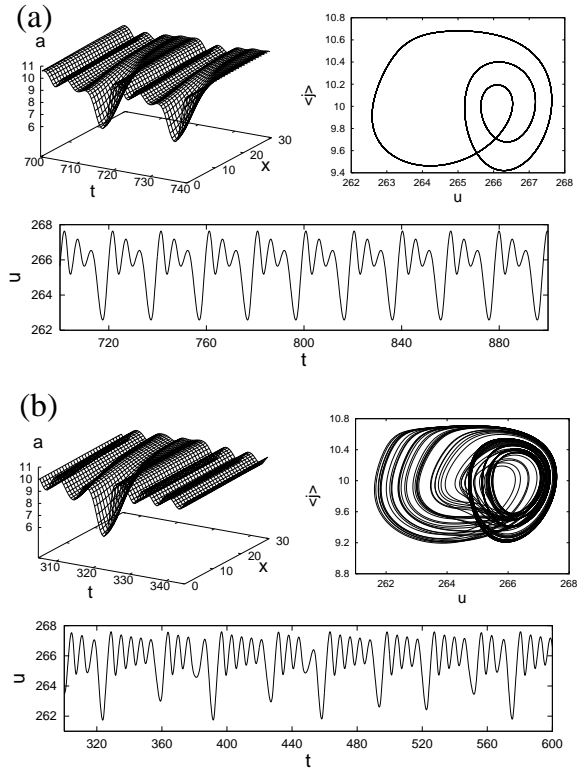,width=8cm}
  \caption{Spatio-temporal spiking patterns:
  as in Fig.~\ref{fig:breathing_scenarios} but (a)  $\eps=13.15$: periodic
    spiking, (b)  $\eps=16.5$: chaotic spiking.}
  \label{fig:spiking_scenarios}
\end{figure}

The filament is stable for $r<0$ at small $\eps$, but
becomes unstable by a Hopf bifurcation for large enough $\eps$.
In this case complex spatio-temporal breathing
(Fig.~\ref{fig:breathing_scenarios})
and spiking patterns (Fig.~\ref{fig:spiking_scenarios})
are expected if the homogeneous fixed point is still stable with respect to
Hopf bifurcations \cite{BOS00}.

\begin{figure}
  \centering
  \epsfig{file=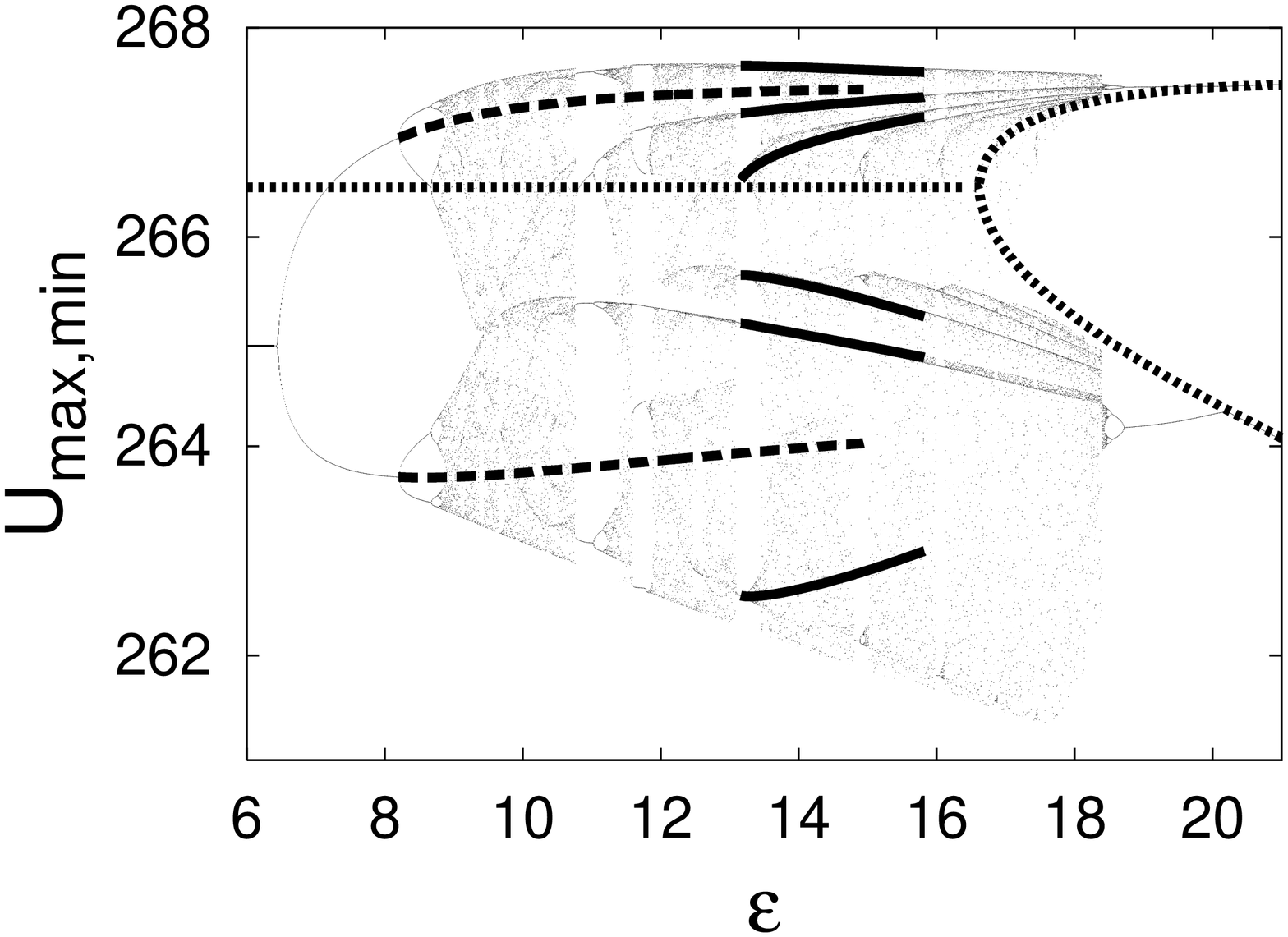,angle=-90,width=7cm}
  \caption{Bifurcation diagram of maxima and minima of the voltage $U$
  versus $\eps$.
  Thick dotted lines: homogeneous solution; thick dashed
  lines: period--one breathing orbit, thick solid lines: period--three
  spiking orbit.
  Parameters and load line as in Fig.~\ref{fig:f_D}(c).}
  \label{fig:bifurcationdiagram}
\end{figure}

Choosing the load line as in Fig.~\ref{fig:f_D}(c) (dotted line), the
homogeneous fixed point on the middle branch is always
unstable against the filamentary mode for $L=30$, but is stable against Hopf
bifurcations up to $\eps < 16.43$ (dotted line in
Fig.~\ref{fig:bifurcationdiagram}). On the other hand, a  Hopf
bifurcation of the half--filament occurs already at $\eps_h^f = 6.4$, leading to 
periodic  filament oscillations 
(Fig.~\ref{fig:breathing_scenarios}(a)) for $\eps > \eps_h^f$. This {\em
  breathing} filament undergoes  a period doubling 
cascade for $\eps > 8.2$ leading to chaos   (c.f.
Fig.~\ref{fig:breathing_scenarios}(b)). We note that the dynamic
behavior is characterized by oscillations of varying amplitude around
the unstable filamentary fixed point, which is typical for a chaotic
breathing scenario.

With further increase of $\eps$ oscillations around the homogeneous fixed
point become important and a competition between the two fixed points
sets in. Thereby the breathing filament transmutes into a 
spiking filament as in Fig.~\ref{fig:spiking_scenarios}(a) for
$\eps=13.15$. Here two oscillations around the homogenous fixed point are
followed by one oscillation around the filamentary fixed point. This
periodic spiking quickly becomes chaotic at further increase of $\eps$,
yielding a chaotic spiking behaviour, which is characterized by a
sequence of almost homogeneous oscillations intermitted by one large
filamentary oscillation (Fig.~\ref{fig:spiking_scenarios}(b)). The
resulting phase portrait is reminiscent of the R\"ossler attractor.
While breathing oscillations are always close to the filamentary fixed
point, the spiking oscillations are related to both the filamentary and
the homogeneous fixed point. The complete bifurcation diagram with respect
to $\eps$ is depicted in Fig.~\ref{fig:bifurcationdiagram}. We note that the
homogeneous limit cycle emerging from a Hopf bifurcation
of the homogeneous fixed point (dotted line) at $\eps = 16.43$ suppresses
chaos with increasing $\eps$ and finally becomes stable against filamentary
oscillations for $\eps>20.2$, and further on determines the dynamics of the system.

\begin{figure}
  \centering
  \epsfig{file=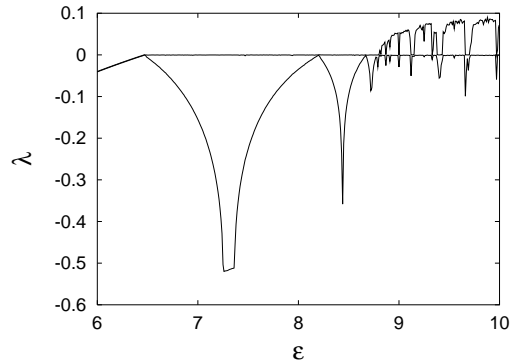,angle=-90,width=7cm}
  \caption{The two largest Lyapunov exponents as a function of the
  bifurcation parameter $\eps$; load line as in
  Fig.~\ref{fig:f_D}} 
  \label{fig:lyapunov}
\end{figure}

Using the Benettin algorithm \cite{BEN80} we have calculated the two
largest Lyapunov exponents $\lambda_1$ and $\lambda_2$ for varying
$\eps$ (Fig.~\ref{fig:lyapunov}). We observe that at most one
Lyapunov exponent is positive, while the second one is zero for a periodic
or chaotic orbit.
Taking into account the third largest Lyapunov
exponent we can then estimate an upper limit for the
fractal dimension of the attractor after Kaplan and Yorke \cite{KAP79}
as $D_{KY} \approx 2.1$. Although the system has an infinite number
of degrees of freedom it is therefore only weakly chaotic. The global
coupling is responsible for the suppresion of extensive spatio
temporal chaos \cite{MEI00}.

\section{Time-delayed feedback control of chaotic breathing and spiking}
\label{sec:control}

We now apply chaos control to stabilize a particular UPO of
period $\tau$ of the uncontrolled system which is embedded in a chaotic
attractor. The control forces $F_a$ and $F_u$ in
eqs. (\ref{eq:a}),(\ref{eq:u}) will be designed such that they do not
suppress chaos by generating a new periodic
solution but only by changing the stability of a solution that already
exists in the uncontrolled system. Hence $F_a$ and $F_u$ vanish exactly
on the target orbit. Such a requirement is in
general met by time--delayed feedback schemes \cite{PYR92}. For
example, we may choose $F_u = F_{\text{vf}}$ with a generic voltage feedback
force defined by 
\begin{equation} \label{ad}
F_{\text{vf}}(t) = u(t)-u(t-\tau) + R F_{\text{vf}}(t-\tau).
\end{equation}
Here $R$ is a memory parameter which damps sudden changes in the
control force by taking into account
multiple delays with a decaying weight of
earlier states of the system
\cite{SOC94}. A control scheme of this type is called extended
time--delay autosynchronization (ETDAS). Such a scheme allows for
  {\em noninvasive} control, since an unstable state may be
  maintained with vanishingly small control forces.

\begin{table}
  \begin{center}
    \begin{tabular}{|l|>{$}c<{$}|>{$}c<{$}|}
      \hline
      Type of control  & F_a & F_u \\
      \hline
      diagonal control  & F_{\text{loc}} & F_{\text{vf}} \\
      global control with voltage feedback  & F_{\text{glo}}& F_{\text{vf}}\\
      global control without voltage feedback  & F_{\text{glo}}& 0\\
      pure voltage control  & 0 & F_{\text{vf}}\\
      local control without voltage feedback & F_{\text{loc}} & 0 \\
      \hline
    \end{tabular}    
  \end{center}
  \caption{Overview of different control schemes with the
  corresponding choices of $F_a$ and $F_u$ }
  \label{tab:control}
\end{table}

For the control force $F_a$ in the spatial degree of freedom we will
alternatively use either a {\em local} control force,
\begin{equation} \label{af}
F_{\text{loc}}(x,t) = a(x,t)- a(x,t-\tau) + R F_{\text{loc}}(x,t-\tau) \quad .
\end{equation}
where every spatial point is controlled independently of its
neighboring points, or a {\em global} control force
\begin{equation} \label{ae}
F_{\text{glo}}(t)= \langle a \rangle(t)- \langle a \rangle (t-\tau) + R
F_{\text{glo}}(t-\tau) \quad .
\end{equation}
with $\langle a \rangle =\frac{1}{L}\int_0^{L} a dx$ 
where the same control force is applied to every spatial point.  This
second choice $F_a=F_{\text{glo}}$ may be experimentally favorable
since the spatial average is related to the total charge in the
quantum well and does not require a spatially resolved measurement.
In the following we will concentrate on the question how the coupling
of the control forces to the spatial and discrete degrees of freedom
influences the performance of the control. The UPO in question will be
a flip orbit, i.e. its largest complex valued Floquet exponent
$\lambda$ will obey $\text{Im} \lambda = \pi/\tau$. This type is practically
important, as it naturally arises in period doubling scenarios, and the
torsion of the orbit associated with $\text{Im} \lambda$ is a necessary
ingredient for time-delayed feedback control to work \cite{JUS97}.
Recently time--delay methods have also been extended to the stabilisation
of orbits without torsion by adding a controller associated with an
additional unstable degree of freedom \cite{PYR01}.

Physically, the control forces may be realized by appropriate electronic
control circuits. $K F_u$ corresponds to an additional control voltage
applied in series with the bias $U_0$, and $K F_a$ may be implemented
by a spatially extended lateral gate electrode which influences the
two-dimensional electron gas in the quantum well locally or globally.

In principle the control performance of time--delayed feedback methods
can be studied by linear stability analysis of the
differential--delay equations (\ref{eq:a}),(\ref{eq:u}) around the
target orbit. This is difficult for a general control scheme,
since the stability of the orbit is governed by Floquet eigenmodes and by the
largest complex valued growth rates $\Lambda$ (Floquet exponent),
which are modified by the control force in a complicated way. Still
there exists a particularly simple control scheme,  called {\sl
  diagonal} control \cite{JUS97,JUS99}, for which  $\Lambda(K,R)$
satisfies the exact equation
\begin{equation}
  \label{eq:characteristic_equation}
  \Lambda+K \frac{1-e^{-\Lambda \tau}}{1-Re^{-\Lambda \tau}} = \lambda.
\end{equation}
where $\lambda$ is the complex Floquet exponent of the uncontrolled
system.
Another control scheme for which analytic results are available is
the {\em Floquet mode control} where the control force is a projection
onto the unstable Floquet mode \cite{BAB02,JUS03}.

In our model we achieve diagonal control by setting $F_a=
F_{\text{loc}}$ and $F_u=F_{\text{vf}}$ (cf.\ Table~\ref{tab:control}),
which is a straightforward extension of the diagonal control for
discrete systems (cf.\ \cite{BLE96}).

\begin{figure}
\begin{center}
\epsfig{figure=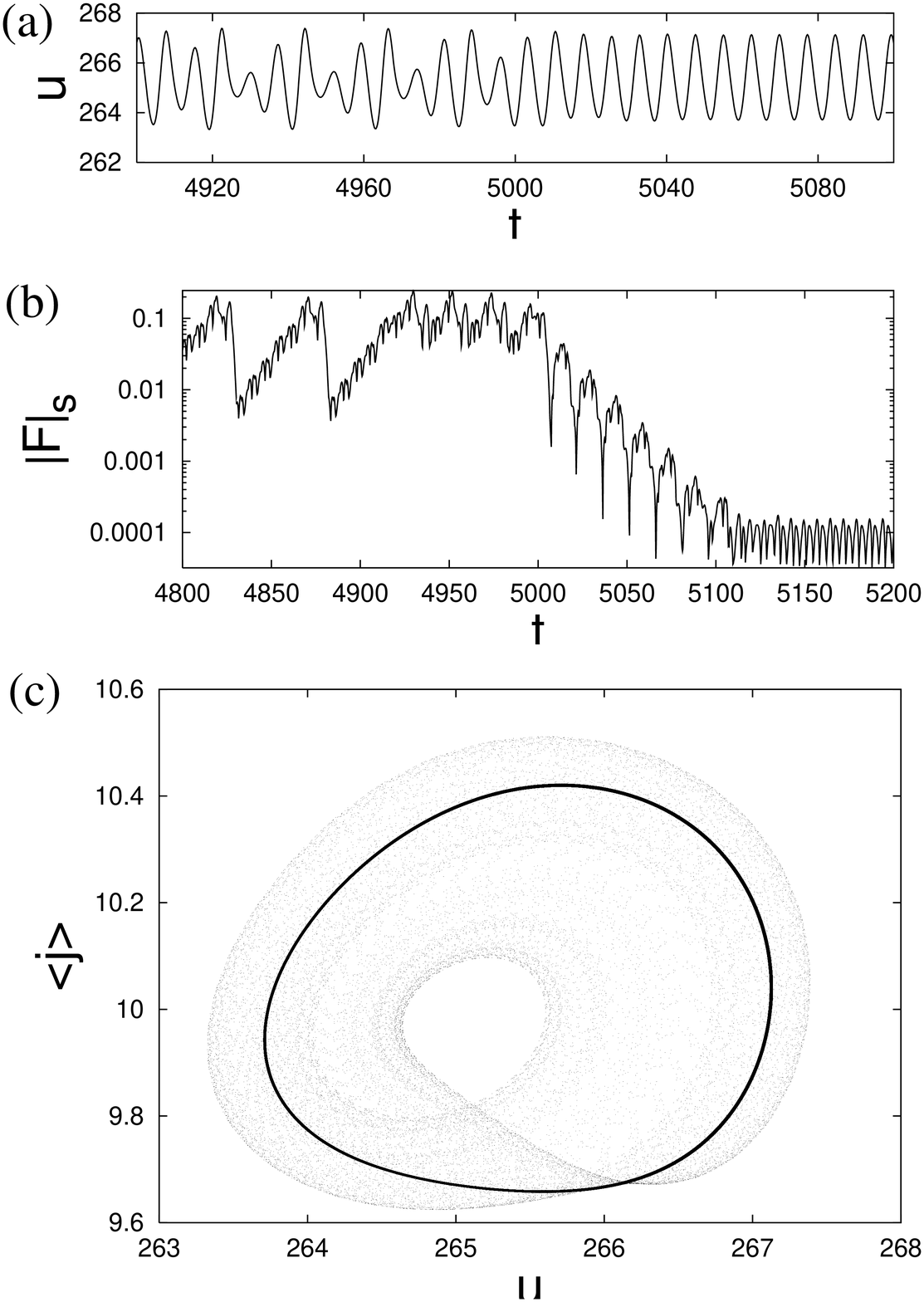,width=80mm}
\end{center}
\caption[]{Diagonal control in the DBRT, where the control force is switched
  on at $t=5000$. (a) Voltage $u$ vs. time, (b) Supremum of the
  control force vs. time, (c) Phase portrait (global current vs.
  voltage) showing the chaotic breathing attractor and the embedded
  stabilized periodic orbit (thick solid curve). Parameters: $r=-35$,
  $\varepsilon=9.1$, $\tau=7.389$, $K=0.137$, $R=0$.
\label{fig:control_breathing}}
\end{figure}

\begin{figure}
\begin{center}
\epsfig{figure=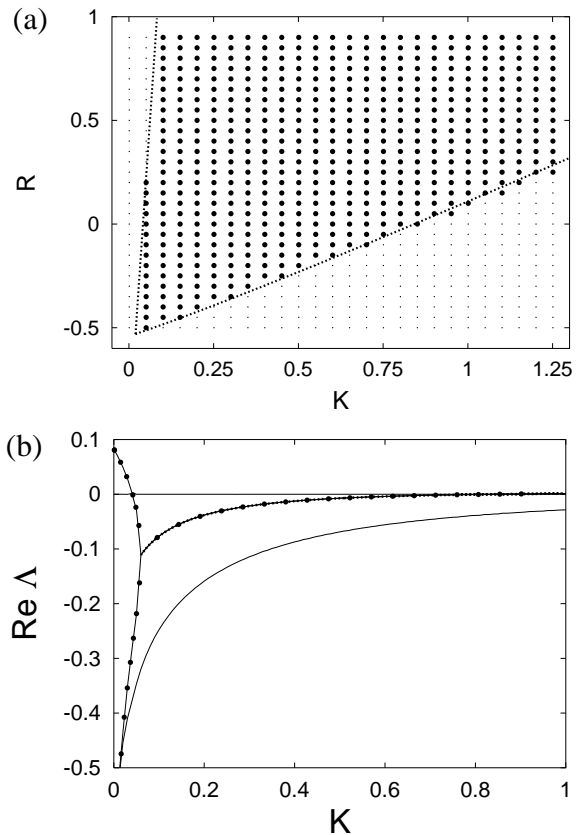,width=80mm}
\end{center}
\caption[]{(a) Control domains in the $K$--$R$ parameter plane for 
diagonal control of the unstable 
periodic orbit of Fig.~4(b) with period $\tau= 7.389$.
$\bullet$ denotes successful
control in the numerical simulation, $\cdot$ denotes no control,
lines: analytical result according to  eq.~(\ref{eq:characteristic_equation}). 
(b) Leading real parts $\Lambda$ of the Floquet spectrum for
diagonal control in dependence on $K$ ($R=0$), $\bullet$ denotes analytical
  results from eq. (\ref{eq:characteristic_equation}).  
\label{fig:diagonal_control}}
\end{figure}

We now want to control the period--one orbit
(dashed line in Fig.~\ref{fig:bifurcationdiagram}) in the chaotic breathing regime
(Fig.~\ref{fig:breathing_scenarios}(b)).  As explained in
Sec.~\ref{sec:scenarios}, this orbit is generated by the supercritical
Hopf bifurcation of the stationary filament at $\eps =
\eps_f^h$. It subsequently becomes unstable by the period doubling
cascade with increasing $\eps$.  From
Fig.~\ref{fig:control_breathing}(b) we note that, before the control is
switched on ($K=0$), a sawtooth--like structure in the (fictitious, since
not yet applied) control force $|F|_s=\text{sup}\{|F_{\text{vf}}|,|F_{\text{loc}}|\}$
appears. This happens as the
uncontrolled system approaches and abandons the UPO ergodically. The
slope of the increasing control force is given by $\lambda$. Such a
pattern is useful for tuning the critical parameter $\tau$
empirically. After the diagonal control is switched on, the control
force decays exponentially as $|F|_s \sim |\exp{\Lambda(K,R)t}|$
(Fig.~\ref{fig:control_breathing}(b)) to a new level which is about
three orders of magnitude smaller than the uncontrolled level. At the
same time the voltage signal becomes periodic
(Fig.~\ref{fig:control_breathing}(a)), and the chaotic attractor in the
phase portrait collapses to a one dimensional periodic orbit.
 
By changing the control parameters $K$ and $R$ we observe that the
regime of successful control in the $K$--$R$ plane (Fig.\ref{fig:diagonal_control}(a))
exhibits a typical triangular shape, bounded by a flip instability (Re
$\Lambda=0$, Im $\Lambda =\pi/\tau$) to its left and by a Hopf (torus)
bifurcation to its lower right. These two boundaries may also be
obtained by solving eq.~(\ref{eq:characteristic_equation}) for
$\text{Re} \Lambda = 0$. We find that the analytical prediction for
the control boundaries are in excellent agreement with the numerical
results (see Fig.~\ref{fig:diagonal_control}(a)).

To confirm the bifurcations at the boundaries we consider the real
part of the Floquet spectrum of the orbit subjected to control for
varying $K$ and $R=0$ (Fig.~\ref{fig:diagonal_control}(b)). Complex conjugate Floquet
exponents show up as doubly degenerate pairs.  The largest nontrivial
exponent decreases with increasing $K$ and collides at negative values
with a branch coming from negative infinity. As a result a complex
conjugate pair develops and the real part increases again. The real part
of the exponent finally crosses the zero axis giving rise to a Hopf
bifurcation. The numerical simulations agree well with the
analytical result.

\begin{figure}
\begin{center}
\epsfig{figure=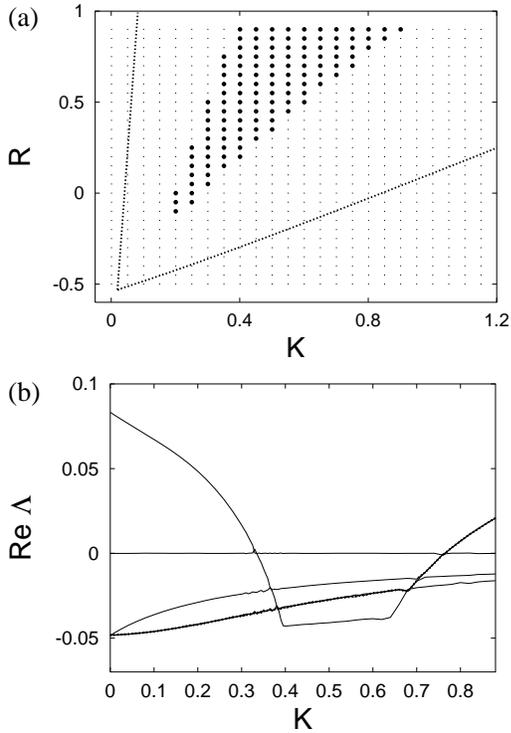,width=70mm}
\end{center}
\caption[]{Same as Fig.~\ref{fig:diagonal_control} for global control
  with voltage feedback ((b): $R=0.7$)
\label{fig:globalwithvoltage}}
\end{figure}

We now replace the local control force $F_a=F_{\text{loc}}$ by the
global control $F_a=F_{\text{glo}}$ without changing the voltage
feedback (i.e. global control with voltage feedback in
table~\ref{tab:control}). Fig.~\ref{fig:globalwithvoltage} shows the
corresponding control regime and Floquet spectrum.  The control domain
looks similar in shape as for diagonal control, although the domain
for the global scheme is drastically reduced.  The shift in the
control boundaries is due to a different scenario in  the Floquet
spectrum. Now $\text{Re} \Lambda$ decreases more slowly with increasing
$K$ than in the diagonal case, and thus the flip bifurcation takes
place at large values of $K$.  At the same time the complex conjugate
pair responsible for the Hopf bifurcation crosses the zero axis with a
larger slope and therefore at a smaller $K$ value than in the diagonal
case.

\begin{figure}
  \centering
\epsfig{figure=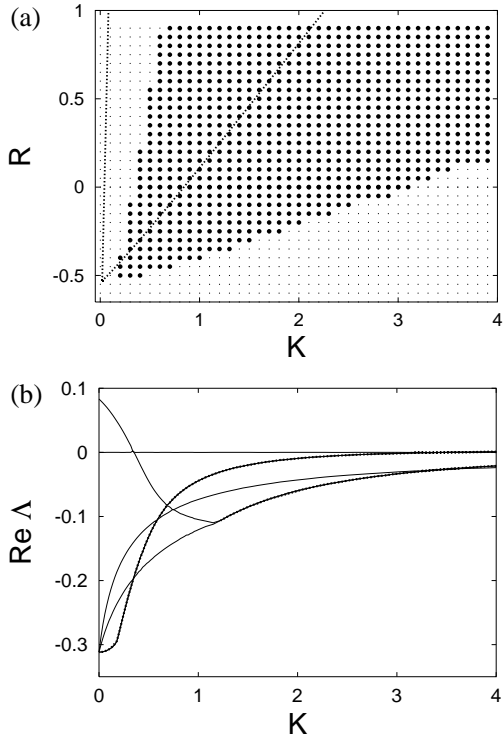,width=70mm}
\caption[]{Same as Fig.~\ref{fig:diagonal_control}
  for global control without voltage feedback ((b): $R=0.1$).
  Note that the scale of the $K$-axis is changed.
  }
  \label{fig:globalwithoutvoltage}
\end{figure}

It is now interesting to note that if we keep $F_a = F_{\text{glo}}$
as before but  remove the voltage feedback completely, 
the control  domain is shifted to higher $K$ values and at the same
time is dramatically increased
(Fig.~\ref{fig:globalwithoutvoltage}(a)). From the Floquet spectrum we see
that after the flip bifurcation the largest Floquet exponent does
not immediately hybridize into a complex conjugate pair, but the Hopf
bifurcation is caused by another complex conjugate pair which is not connected
to the largest Floquet exponent. Thereby the Hopf bifurcation is
suppressed and the control regime is increased. This behavior is very
similar to the one observed in a different reaction--diffusion model
(modeling a heterostructure hot electron diode, HHED) \cite{BEC02}, where it
was found that additional control of the global variable $u$ may
gradually reduce the control regime to zero.

\begin{figure}
  \centering
\epsfig{figure=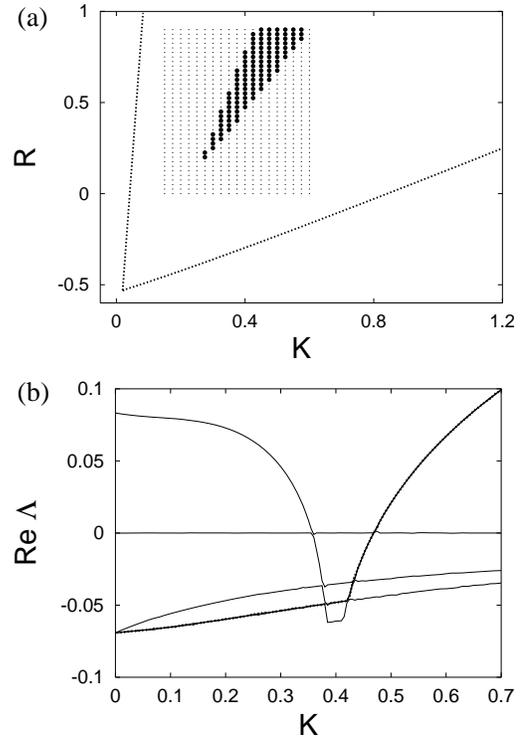,width=70mm}
\caption[]{Same as Fig.~\ref{fig:diagonal_control} for pure voltage control
((b): $R=0.6$)}
  \label{fig:purevoltage}
\end{figure}

From the practical point of view the most relevant control scheme is
the pure voltage control, i.e. $F_u=F_{\text{vf}}$, $F_a=0$,
  since the voltage variable may be conveniently measured and
  manipulated by an external electronic device. The corresponding
control domain and Floquet exponents are shown in
Fig.~\ref{fig:purevoltage}. Here the control regime is even somewhat
smaller than in the case of global control with voltage feedback but the shape
of the control regime and the Floquet spectrum are qualitatively very
similar. It is encouraging that this kind of control is at all
possible, since it opens up the opportunity to conveniently
study chaos control in a real world device.

\begin{figure}
  \centering
\epsfig{figure=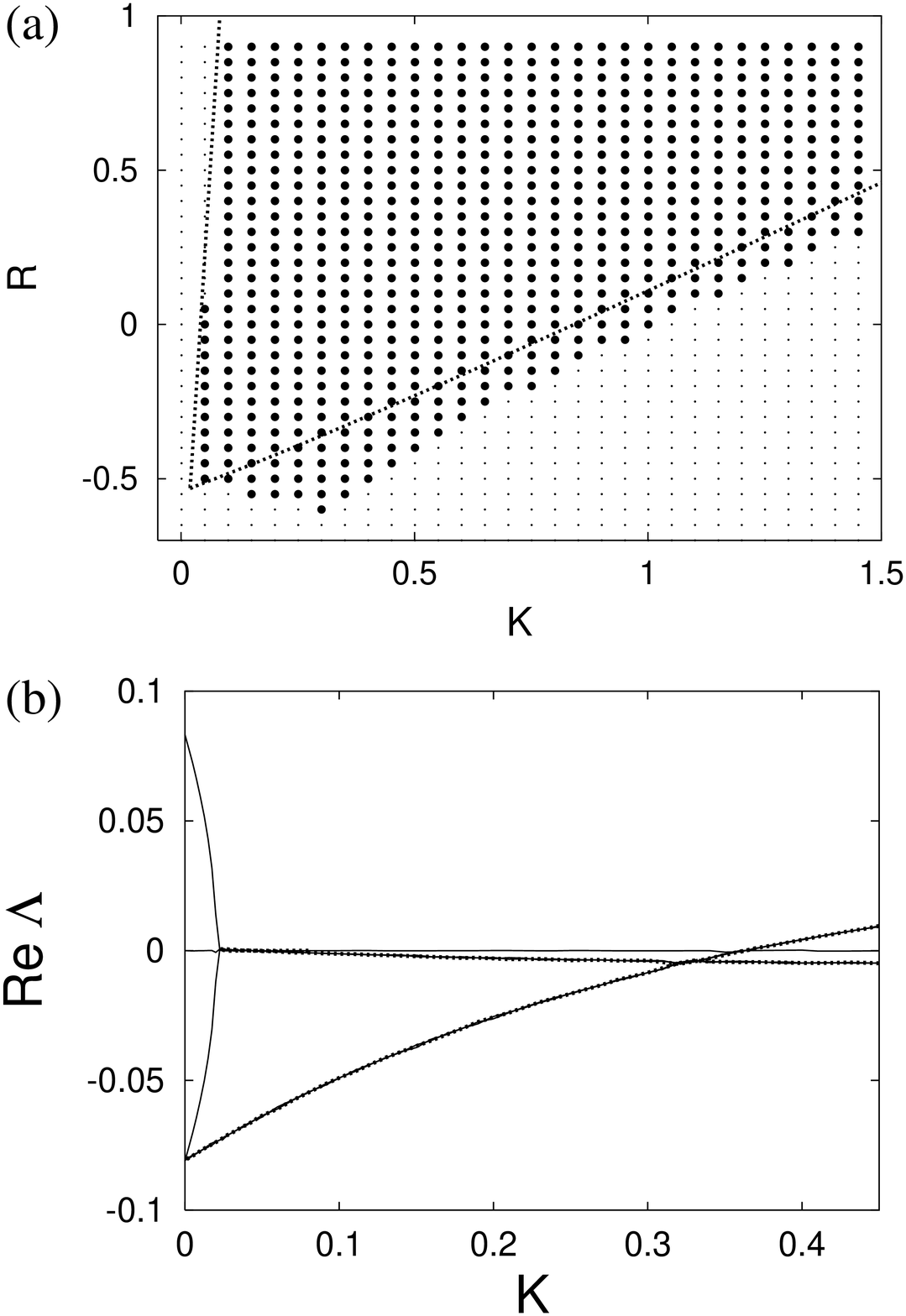,width=70mm}
\caption{Same as Fig.~\ref{fig:diagonal_control} for local control without
voltage feed back ((b): $R=-0.55$).}
  \label{fig:localwithoutvoltage}
\end{figure}

We finally consider the case of local control without voltage feedback
(Fig.~\ref{fig:localwithoutvoltage}). Here the control regime is
surprisingly even larger than for diagonal control. The shape of the
control regime is not triangular any more as before, but has an additional
edge at low $K$ and $R$ values. The reason for this edge can be
explained from the Floquet diagram. Here a Floquet exponent from
below collides with the largest Floquet exponent at positive real values
before the flip bifurcation occurs.  This complex conjugate pair then
crosses the zero axes from above and undergoes an inverse Hopf
bifurcation. For larger values of $K$ another complex conjugate pair
performs a second Hopf bifurcation seemingly unrelated to the first
one. 

An overview of the different coupling schemes considered in this
section is given in
table~\ref{tab:control}. By comparing the different control schemes,
we may characterize the influence of the voltage control. If the
voltage control is switched on, the largest Floquet exponent rises
quickly after the collision with an exponent coming from below, and
participates in the Hopf bifurcation at the right boundary of the
control regime. If on the other hand the voltage control is switched
off, the largest Floquet exponent only rises slowly after the
collision, and often the Hopf bifurcation at the right boundary is
caused by a complex conjugate pair which is indepent of the
largest Floquet exponent. In this case the Hopf boundary is shifted to
larger values of $K$ than for diagonal control. The choice of the
control force $F_a$ influences the decrease of $\text{Re} \Lambda(K)$
at small $K$. For $F_a = F_{\text{loc}}$ this decrease is large and the
flip boundary practically coincides with the one obtained for diagonal
control, while for $F_a = F_{\text{glo}}$ the flip boundary is shifted
to higher $K$ values since $|\partial_K\text{Re} \Lambda(K)|$ is
small. For $F_a=0$ the slope $|\partial_K\text{Re} \Lambda(K)|$ is even smaller,
shifting the boundary to even higher $K$ values.

\begin{figure}
  \centering
  \epsfig{figure=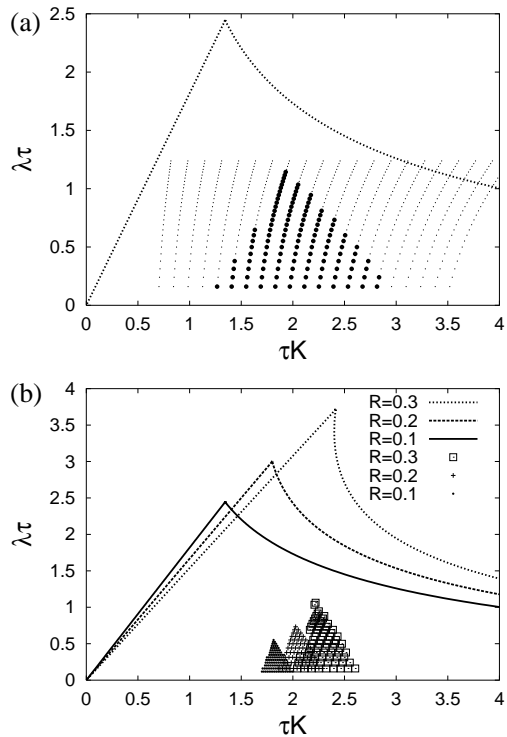,width=70mm}
  \caption{Control domain in the $\lambda-K$ plane (a) for global control
          with voltage feedback ($R=0.1$), $\bullet$: simulation;
          dotted line: analytical result for diagonal control from
          eq.~(\ref{eq:characteristic_equation}), (b) pure voltage control
          (symbols: simulation with different values of $R$ as given in the
          legend; lines: analytical results for diagonal control from
          eq.~(\ref{eq:characteristic_equation})).
          }
  \label{fig:lamdak}
\end{figure}

The period-one orbit can be controlled continuously in a whole range of values
of $\eps$ (cf. thick dashed lines in Fig.~\ref{fig:bifurcationdiagram}
for diagonal control).
Note that since the period $\tau$ of the UPO depends on $\eps$, $\tau$ needs
to be readjusted while sweeping $\eps$. It is then even
possible to stabilize the period-one orbit in higher periodic windows, where
the target orbit is obviously not part of the attractor. This opens up the
possibility of obtaining stable
self-sustained voltage oscillations independently of parameter
fluctuations. The Floquet exponent $\lambda$ also depends on $\eps$, and
this allows us to study the control 
performance as a function of the largest Floquet exponent of the UPO.
In Fig.~\ref{fig:lamdak}(a) the control performance for global control
with voltage feedback for fixed $R$ is plotted. As expected, the regime of
control is considerably smaller than the analytical predictions
for diagonal control. Fig.~\ref{fig:lamdak}(b) shows the control
performance for pure voltage feedback and different values for $R$.
Although the control regimes are much smaller than predicted for 
diagonal control, the trend in the shift of the control
regime for increasing $R$ is qualitatively similar. 

\begin{figure}
  \centering
  \epsfig{figure=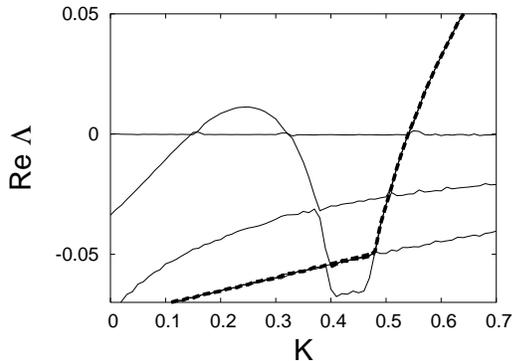,angle=-90,width=70mm}
  \caption{ Leading real parts $\Lambda$ of the Floquet spectrum for
  pure voltage control of a stable period-one breathing orbit in dependence
  on $K$ ($R=0.6, \eps=8.0$) }
  \label{fig:reentrance}
\end{figure}

Surprisingly we find a finite $K$ value for the flip boundary as
$\lambda \tau \to 0$ (Fig.~\ref{fig:lamdak}(b)). This is in marked  contrast to
the analytical results for diagonal control, where the flip
boundary  extrapolates to zero for $\lambda \tau \to 0$. To understand why this
is the case, let us consider a situation where we apply pure
voltage control to a {\em stable} orbit. From the Floquet spectrum in 
Fig.~\ref{fig:reentrance} we see that with increasing control force
the stable orbit is destabilized by a flip bifurcation, then becomes
stable again by a second flip bifurcation, before it finally
undergoes the usual Hopf bifurcation at high $K$ values. Such a 
reentrance scenario does not happen for diagonal control. 

\begin{figure}
  \centering
  \epsfig{figure=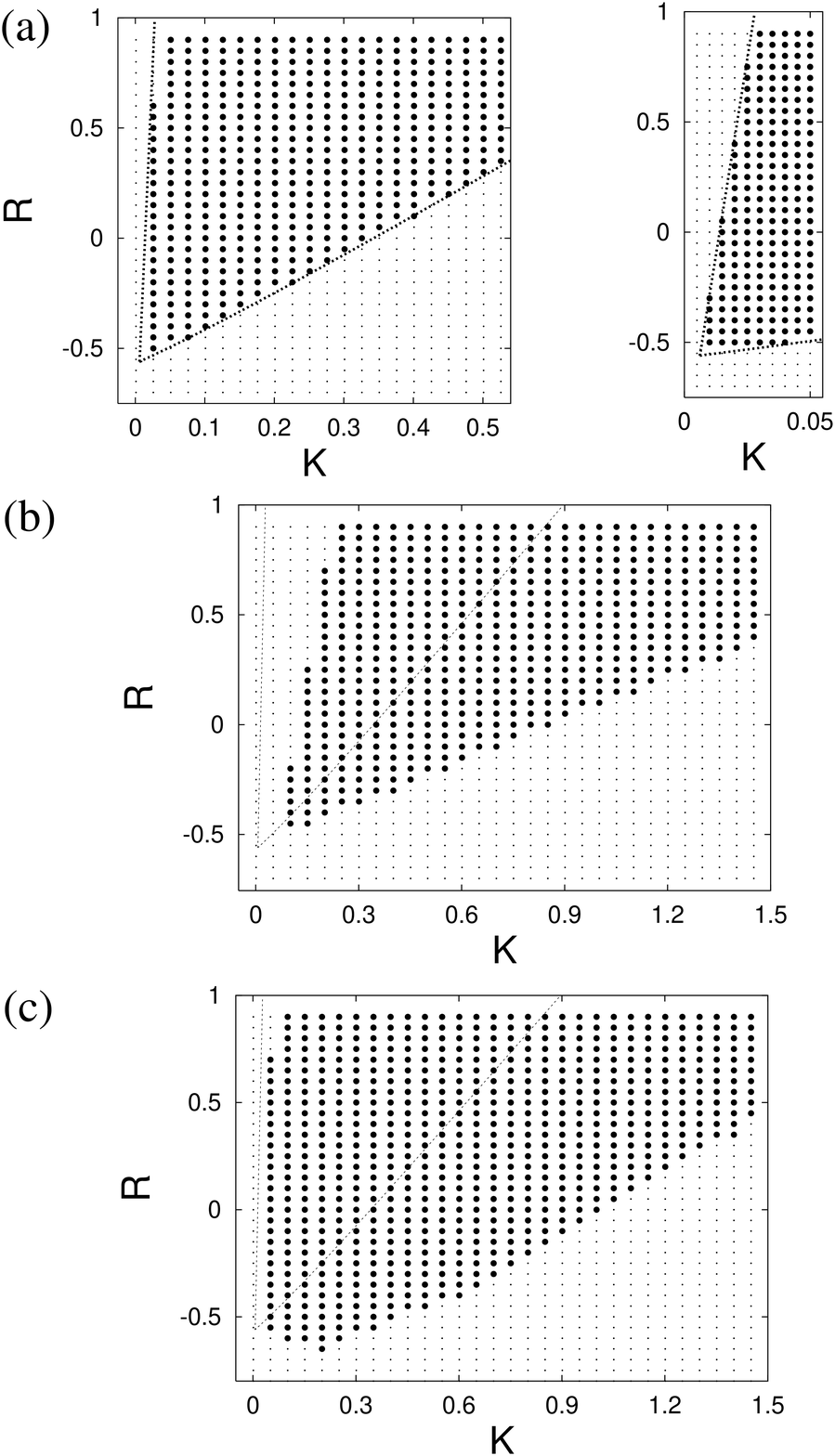,height=80mm}
  \caption{Control domains for spiking orbit. Lines: analytical results
  for diagonal control. (a) diagonal control, (b) global control
  without voltage feedback, (c) local control without voltage feedback
(Parameters: $\eps = 13.44$, $\tau =20.15$)}
  \label{fig:spiking_control}
\end{figure}

So far we have stabilized the period-one breathing orbit. We have
investigated whether the obtained results are specific 
for a given UPO, or whether similar results can be 
observed for different orbits such as the period-three spiking orbit
(cf. thick solid lines in Figs.~\ref{fig:spiking_scenarios},\ref{fig:bifurcationdiagram})
which stands for the second basic dynamical pattern (spiking) in the DBRT
system. Fig.~\ref{fig:spiking_control} shows the control regimes for
different control schemes. Again we note that diagonal control is very
well predictable by the analytic formula
(\ref{eq:characteristic_equation})
(Fig.~\ref{fig:spiking_control}(a)), whereas omitting the voltage control
shifts the Hopf border to large $K$ values, and replacing local with
global control shifts the flip boundary to larger $K$ values. For pure
  voltage control and global control with voltage feedback no domain of
  control was found. Nevertheless features of the different control schemes
  are similar for both breathing and spiking orbits.

\section{Conclusions}

We have applied different schemes of chaos control by time--delay
autosynchronization to a globally coupled reaction--diffusion model
describing charge transport in a semiconductor nanostructure, viz.
a double-barrier resonant tunneling diode (DBRT).
The spatio--temporal dynamics and bifurcation scenarios
of the uncontrolled DBRT in an external circuit display a variaty of complex
patterns, including breathing and spiking oscillations either of which can be
periodic or chaotic.  

We have shown that by using time--delayed feedback methods it is possible to
stabilize unstable breathing or spiking patterns.
Different control scheme were compared with respect to efficiency,
and quantitative comparisons of their respective control domains in the
$R-K$ parameter space were given and
interpreted in terms of the Floquet spectra. Those
turned out to be helpful to gain insight into the control mechanisms
and understand why the control performance may, e.g., be improved by
omitting control of the voltage variable.
As a control scheme which is particularly simple to realize experimentally,
albeit with a small control regime which requires careful adjustment of
the control parameters $R$ and $K$, we have singled out pure voltage control.
Our findings may be important for obtaining stable
self-sustained voltage oscillations in resonant tunneling diodes
independently of parameter fluctuations.

This work was supported by DFG in the framework of Sfb 555.

\appendix
\section{}
\label{sec:app}
In this Appendix we provide analytical expressions for the functions
$D(a)$ and $f(a,u)$ appearing in eq.~(\ref{eq:a}) and relate the
dimensionless quantities used throughout this paper to their respective
dimensional physical quantities marked by a tilda. 

The voltage drop across the DBRT $\tilde{u}(t)$ and the external bias voltage
$\tilde{U}_0$ are related to their dimensionless counterparts by
$u=e\tilde{u}/(k_BT)$ and $U_0=e\tilde{U}_0/(k_BT)$, respectively, with
the temperature $T$, the electron charge $e<0$, and Boltzmann's constant
$k_B$. The two--dimensional electron density is rescaled via $a =
\tilde{a}\pi \hbar^2 /(m k_B T)$, where $m$ is the effective mass of the
electron in the well. Time and space coordinates are rescaled by
$t=\tilde{t} \Gamma_L/\hbar$ and $x = \tilde{x}/\sqrt{\hbar \mu k_B T
  /(-e \Gamma_L)}$ where $\Gamma_L=\Gamma_R$ is the transition rate for
electrons from the emitter to the quantum well (and from the quantum well
to the collector) and $\mu$ is the electron
mobility. 

This yields a rescaling of current and resistance as
$j=\tilde{\jmath}\pi \hbar^3/(em k_B T \Gamma_L)$ and $r = \tilde{r}e^2 m A
\Gamma_L/(\pi \hbar^3)$, respectively, where $A$ is the cross sectional area
  of the device. Note that the effective
resistance in general contains two terms $\tilde{r}=R-k$, where $R$
represents an external resistor and $k$ arises from an active circuit
with a voltage gain $U_c = k \tilde{\jmath}A$ \cite{MAR94}. In the case $k>R$ this leads
to $r<0$ with particularly interesting chaotic scenarios, as shown in
the main text. The time-scale ratio is given by $\varepsilon=RC\Gamma_L/\hbar$,
where $C$ is the total capacitance of the circuit.

The effective diffusion coefficient $D(a)$ may be calculated by
considering  a generalized form of Einstein's relation 
which covers the full range from Fermi-degenerate to nondegenerate  
conditions and a drift term resulting from the change in the local
potential due to Poisson's equation \cite{CHE00}:
\begin{eqnarray} 
D(a) = a
\left(\frac{d}{r_B} +\frac{1} {1-\exp\left[-a\right]}\right),
\label{eq:D_a}
\end{eqnarray}
where $r_B \equiv (4\pi\epsilon\epsilon_0 {\hbar}^2)/(e^2 m)$ is the
effective Bohr radius in the semiconductor material, $\epsilon$ and $\epsilon_0$ are the
relative and absolute permittivity of the material, respectively,
and $d$ is the effective thickness of the double-barrier structure. 

The function $f(a,u)$ is obtained from a microscopic consideration of
the tunneling currents  $J_{ew}(a,u)$ and $J_{wc}(a)$ from the emitter
to the quantum well and from the well to the collector, respectively
\cite{SCH02}:
  \begin{gather}
    f(a,u)=\left[\frac{1}{2} + \frac{1}{\pi}
      \arctan{\left(\frac{2}{\gamma}(x_0-\frac{u}{2} +
          \frac{d}{r_B}a)\right)}\right]
    \nonumber \\
    \times\left[\ln {\left(1+\exp{(\eta_e-x_0 + \frac{u}{2} -
            \frac{d}{r_B}a)}\right)} -a \right]-a
   \end{gather}
Here $\gamma$ and $x_0$ describe the broadening and the energy level of  the electron
states in the quantum well, and $\eta_e$ is the dimensionless Fermi level
in the emitter, all in units of $k_BT$. In this article we choose $\gamma=6$, $d/r_B=2$, $\eta_e=28$
and $x_0 = 114$. Typical physical values correspond to units of time, space,
voltage, electron density, and current density
of the order of $3.3$ ps, $100$ nm, $0.35$ mV, $10^{10} \text{cm}^{-2}$,
$500 \text{ A} \text{cm}^{-2}$, respectively, for $d=20 \text{ nm}$,
$\Gamma_L=\Gamma_R=0.2 \text{ meV}$,
$k_BT=0.32 \text{ meV}$ ($T=4 \text{ K}$ ).

\bibliographystyle{prsty-fullauthor}

\end{document}